\documentclass[a4paper,11pt]{article}
\usepackage{pos}
\usepackage{subfigure}
\usepackage{graphicx}
\usepackage{setspace}
\usepackage{sidecap}
\usepackage{wrapfig}

\title{Status of the Trinity PeV Neutrino Observatory}

\author*[a]{Sofia Stepanoff}
\onbehalf{for the \emph{Trinity} Collaboration}

\affiliation[a]{School of Physics and Center for Relativistic Astrophysics, Georgia Institute of Technology,\\
  Atlanta, Georgia, USA}
\emailAdd{sofiastepanoff@gatech.edu}

\abstract{The Trinity Neutrino Observatory aims to detect tau neutrinos in the energy range of 1\,PeV to 10\,EeV. We are developing the observatory in three stages. The first stage, known as the Trinity Demonstrator, was deployed in Fall 2023. The Demonstrator serves as a pathfinder for the full observatory and will inform the design of the first Trinity Telescope. We discuss the status and initial results of the Trinity Demonstrator. In 346 hours of observations with the Demonstrator, we do not identify a neutrino candidate event.}

\ConferenceLogo{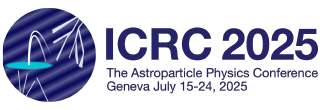}

\FullConference{39th International Cosmic Ray Conference (ICRC2025)\\
 15–24 July 2025\\
Geneva, Switzerland\\}

\begin{document}
\maketitle
\newcommand{\trinitydem}{\textit{Trinity} Demonstrator}
\newcommand{\trinityone}{\textit{Trinity} One}
\newcommand{\trinityobs}{\textit{Trinity Neutrino} Observatory}
\newcommand{\agn}{active galactic nuclei (AGN)}
\newcommand{\ic}{\textit{IceCube}}
\newcommand{\ictwo}{\textit{IceCube-Gen2}}
\newcommand{\ngc}{NGC1068}
\newcommand{\txs}{TXS 0506+056}
\newcommand{\pgamma}{$p\gamma$}
\newcommand{\pp}{$pp$}
\newcommand{\ev}{eV}
\newcommand{\gev}{GeV}
\newcommand{\pev}{PeV}
\newcommand{\eev}{EeV}
\newcommand{\tev}{TeV}
\newcommand{\mev}{MeV}
\newcommand{\nt}{$\nu_\tau$}
\newcommand{\nel}{$\nu_e$}
\newcommand{\nm}{$\nu_\mu$}

\section{Introduction}

High-energy neutrinos (HE, 1\,TeV - 1\,PeV) have proven to be versatile probes to address questions ranging from understanding cosmic-ray accelerators \cite{Kurahashi2022}, neutrino particle physics \cite{Aartsen2018a}, and new physics beyond the standard model \cite{Ellis2018}. We expect that observations of very-high energy (VHE, 1\,PeV-1\,EeV) neutrinos will be of similar impact. VHE and higher energy neutrinos provide clues about the origin of ultra-high-energy cosmic rays (UHECRs) and the evolution of their sources \cite{Aloisio2015}, and probe neutrino physics at energies unachievable with particle accelerators. 

However, observations of VHE neutrinos are rare. A handful of neutrinos have been detected at PeV energies, and one stunning 220\,PeV event was reported by the KM3NET team \cite{Aiello2025ObservationKM3NeT}. Detecting VHE neutrinos requires extraordinarily large detector volumes to compensate for the steeply falling neutrino fluxes. Several efforts are underway to explore different techniques that have the potential to overcome these challenges and open the VHE neutrino sky by utilizing the Earth, atmosphere, water, or ice as a neutrino target, and employing Cherenkov light detectors, scintillators, or radio detectors to measure the neutrino interaction. Some proposed and ongoing experiments are Beacon \cite{Southall2023}, GRAND \cite{GRANDCollaboration2018}, RNO-G \cite{Aguilar2020DesignRNO-G}, IceCube Gen2 \cite{IceCubeGen22020}, and \emph{Trinity}. Each approach has its strengths and weaknesses, is sensitive in different energy bands, and views different parts of the sky. Achieving all-sky coverage with uniform sensitivity across a broad energy range requires several operating instruments. 

We pursue the \emph{Trinity} Neutrino Observatory, three arrays of imaging atmospheric Cherenkov telescopes, on mountain tops aimed at the horizon to detect Earth-skimming VHE tau neutrinos. Here, we provide a status update of the \emph{Trinity} project. We begin in section \ref{sec:observatory} with an overview of the concept behind \emph{Trinity}, and the conceptual design of the \emph{Trinity} Neutrino Observatory. The development of \emph{Trinity} is staged in three phases. The first phase, discussed in section \ref{sec:demonstrator}, consisted of the development and operation of the Demonstrator to inform the design of the \emph{Trinity} telescopes and demonstrate the concept. Results of the Demonstrator are presented in section \ref{results}.

\begin{SCfigure}[0.7][tb]
\centering
\includegraphics[angle=0,width=0.7\columnwidth]{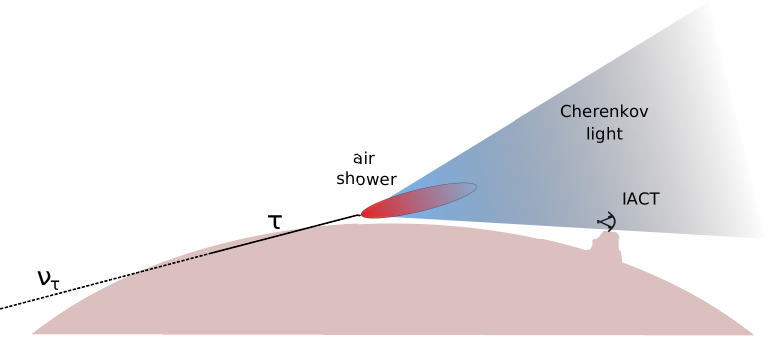}
\caption{
\emph{Trinity} detection concept. Following a tau-neutrino interaction inside the Earth, the resulting tau decays in the air, starting a particle shower imaged with a \emph{Trinity} Cherenkov telescope.}
\label{fig:EarthSkimming}
\vspace{-2ex}
\end{SCfigure}
\vspace{-3mm}
\section{The \trinityobs\ Concept}\label{sec:observatory}

\emph{Trinity} detects Earth-skimming VHE tau neutrinos, which are neutrinos entering the Earth under a shallow angle of $<10^\circ$ (see Figure \ref{fig:EarthSkimming}). The length of the neutrino trajectory resulting from the shallow entrance angle, combined with the high energy of the neutrino, yields a high probability for the neutrino to undergo a charged-current interaction and generate a tau that emerges from the ground and decays in the atmosphere. The tau decay initiates an extensive particle shower, emitting Cherenkov emission that can be collected by Cherenkov telescopes from as far away as 200\,km \cite{otte_studies_2019}.

The idea to detect Earth-skimming neutrinos with Cherenkov telescopes has been proposed by \cite{Fargion2006} and explored by ASHRA \cite{Ogawa2019} and MAGIC \cite{gora_magic_2017}. While the ASHRA team has pointed its telescope at a relatively nearby mountain, MAGIC has pointed its telescopes at the horizon, resulting in a larger effective area for VHE neutrinos. \emph{Trinity} follows the MAGIC approach by pointing the telescope slightly below the horizon. \emph{Trinity} further maximizes its effective area by using silicon photomultipliers, which are more red sensitive than MAGIC's photomultiplier tubes and can, therefore, detect Cherenkov light from farther distances \cite{otte_studies_2019}.

\begin{SCfigure}[1.0][tb]
\centering
\includegraphics[angle=0,width=0.7\textwidth]{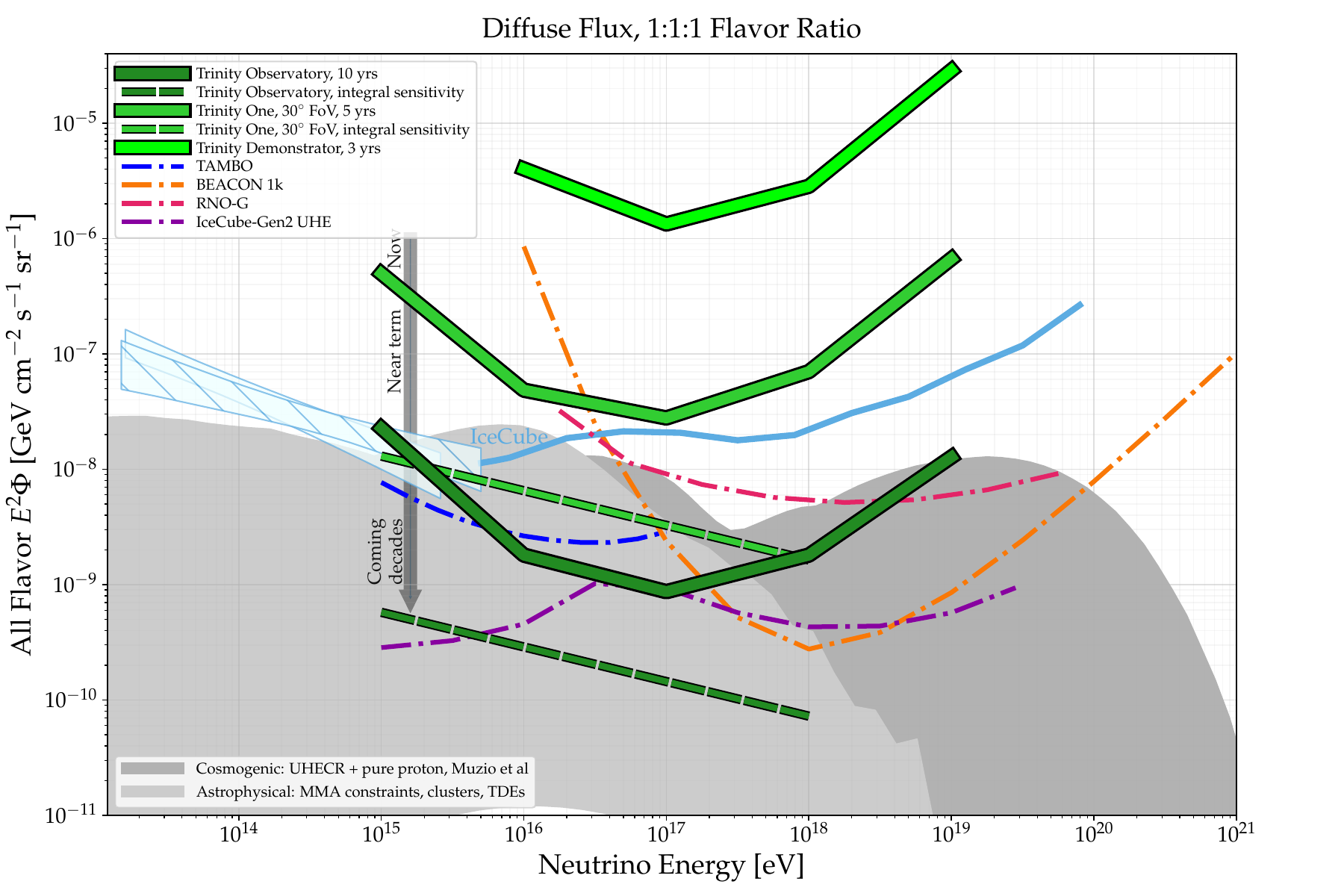}
\caption{\footnotesize  Differential flux sensitivity (90\% C.L.\ upper limit) of the  \emph{Trinity} Demonstrator, \emph{Trinity} One, and the \emph{Trinity} Neutrino Observatory. The sensitivity calculations assume a 20\% duty cycle. The integral detection sensitivity for \emph{Trinity} One is calculated under the assumption that the astrophysical flux (light blue hashed bow tie) extends like a power law. The light gray and dark gray areas outline models for astrophysical and cosmogenic neutrino fluxes, respectively. Also shown is  the constraint on $E_\nu>$ \pev\ neutrino emission by \ic. Figure adapted from \cite{Ackermann_Bustamante_Lu_Otte_Reno_Wissel_Ackermann_Agarwalla_Alvarez-Muñiz_AlvesBatista_etal._2022}.}
\label{fig:senstivity}
\vspace{-4ex}
\end{SCfigure}

Trinity's design sensitivity for diffuse neutrino fluxes, \emph{i.e.}, astrophysical and cosmogenic neutrinos, shown in Figure \ref{fig:senstivity} is accomplished with 18 wide-angle Cherenkov telescopes. Each telescope has a $60^\circ$ horizontal field of view, 20 times larger than one MAGIC telescope \cite{Cortina2016}. The 18 telescopes are distributed on at least three sites, each telescope viewing an independent volume of atmosphere. The diffuse-flux sensitivity of \emph{Trinity}, thus, scales linearly with time and the combined horizontal field of view of all \emph{Trinity} telescopes, which explains why, after ten years, the diffuse-flux sensitivity of the \emph{Trinity} Neutrino Observatory will be 36 times the sensitivity of a single telescope after five years (\emph{Trinity} One). The \emph{Trinity} Neutrino Observatory will overlap at the low-end of its accessible energy range with IceCube and KM3NeT and at the upper-end with radio-based neutrino experiments like Beacon. The project cost for the final observatory is estimated at around $20$\,million USD, \emph{Trinity} presents itself as a cost-effective solution to cover the VHE energy range.


\section{The \trinitydem\ Telescope}
\label{sec:demonstrator}

The first phase of developing the \emph{Trinity} Neutrino Observatory involves building and operating the \emph{Trinity} Demonstrator. The purpose of the Demonstrator is to validate the \emph{Trinity} concept, demonstrate that remotely operating is feasible, and build the expertise necessary before developing the next phase, \emph{Trinity} One. Among the most relevant questions addressed by the Demonstrator is the question about sources of background events that could be misidentified as neutrino events.

\begin{wrapfigure}[18]{l}{0.4\textwidth}
\includegraphics[angle=0,width=0.4\columnwidth]{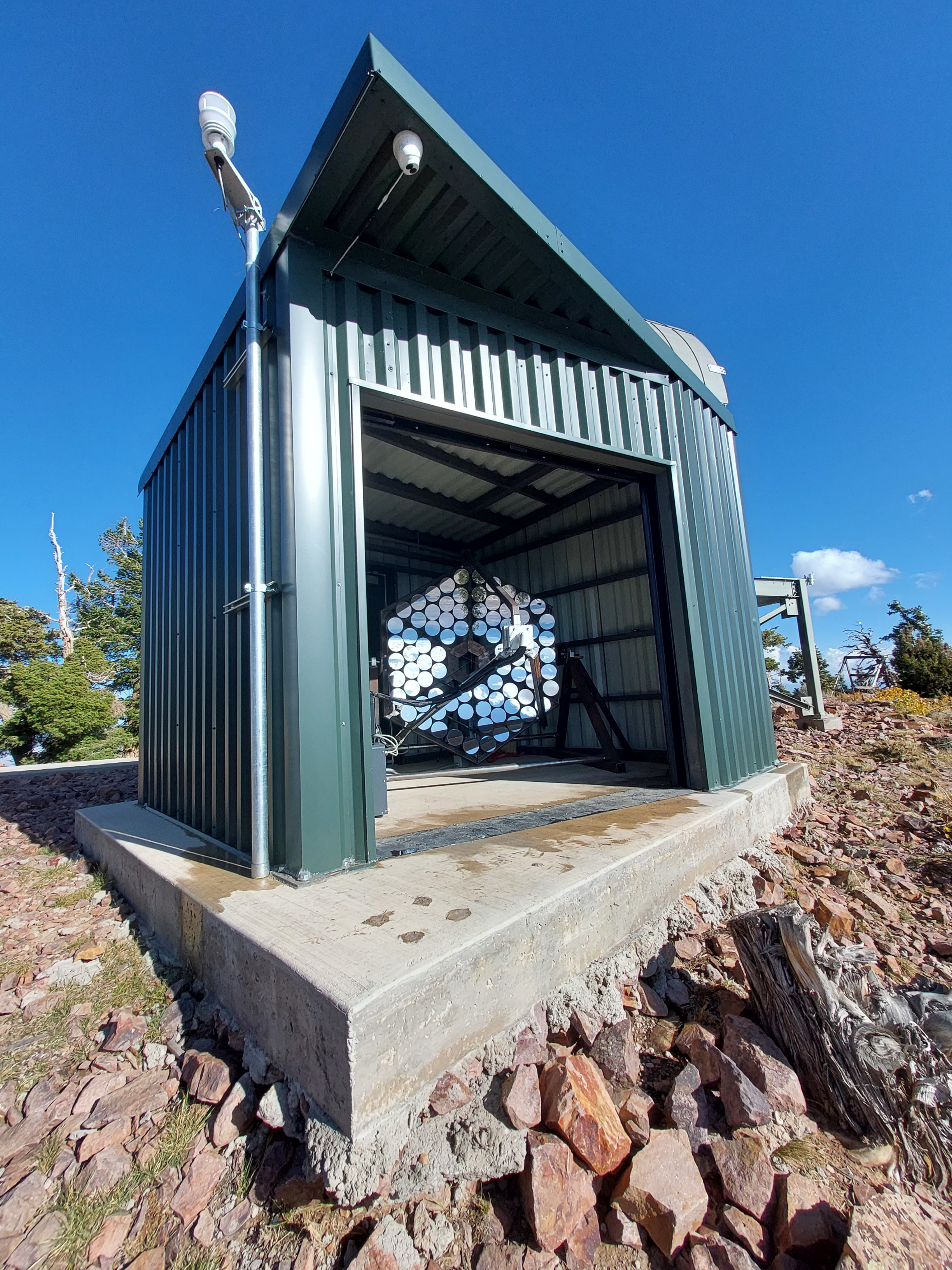}
\caption{\footnotesize The \emph{Trinity} Demonstrator inside its building on Frisco Peak, Utah.}
\label{fig:demonstrator}
\vspace{-2ex}
\end{wrapfigure}

Figure \ref{fig:demonstrator}, shows the telescope deployed on Frisco Peak, Utah at an elevation of 2930\,m. The telescope uses a conventional Davies-Cotton optics realized with a tessellated $1.36$\,m$^2$ light collection surface consisting of 77 15\,cm diameter circular mirror facets. Located in the focal plane of the telescope is a 256-pixel silicon photomultiplier (SiPM) camera yielding a $0.24^\circ$ angular resolution and a $3.8^\circ\times3.8^\circ$ field of view. The SiPMs are S14161-6050HS-04 devices from Hamamatsu. The SiPM signals are conditioned with eMUSIC ASICs \cite{Gomez2016} and recorded by an AGET digitizer with a sampling speed of 100\,MS/s and 12\,bit resolution \cite{Pollacco2018}. The camera readout is triggered and all camera signals digitized whenever the signal in one SiPM is above the equivalent of 20\,photoelectrons. The stability of the camera and the readout is monitored during data taking with a fast LED light flasher producing a 10-nanosecond-lasting light flash once every second. The telescope points below the horizon, such that $0.7^\circ$ of the $3.8^\circ$ vertical camera field of view images the sky above the horizon, and $3.1^\circ$ images the ground below the horizon. The top part of the camera serves as a veto for down-going air-showers from cosmic rays.

The construction of the Demonstrator was completed in Fall 2023, and commissioning ended in Summer 2024. Details about the commissioning of the Demonstrator, its performance, and calibration can be found here \cite{bagheri_commissioning_2025}.

Here we present the first results from 346 hours of data recorded with the Demonstrator between October 2024 and May 2025, requiring cloud bases at least 300\,m above the Demonstrator's elevation. During that time, the readout triggered 979,957 times. 
\vspace{-4mm}
\section{Demonstrator Data Analysis and Results \label{results}} 

For the data analysis, we follow the steps of a conventional analysis of Cherenkov telescope data. Given the Demonstrator's sensitivity, we do not expect to record events from neutrino-induced air-shower signals. The main goal of our analysis is to identify potential sources of background and demonstrate the ability to suppress them.

The analysis of a recorded event starts by extracting the signals in each camera pixel. The extracted signals are corrected for gain differences relative to the camera mean using interleaved LED flashes that uniformly illuminate the camera \cite{bagheri_commissioning_2025}. The thus flatfielded signals are then calibrated in photoelectrons, \emph{i.e.}, detected photons. The calibration factor for signal to photoelectrons have been measured during a Demonstrator site visit and are corrected for temperature drifts of the SiPM breakdown voltage during the analysis \cite{bagheri_commissioning_2025}.

Following the calibration of the camera signals, the camera image of the event is then cleaned by suppressing pixels that only contain noise signals. The process, dubbed image cleaning, starts by identifying the brightest pixel in the camera sector, triggering the event. In the second step, all core pixels are identified, \emph{i.e.}, all pixels with a signal similar to that of the brightest pixel. Of these, all core pixels that touch at least one other core pixel with a common side or corner are retained. In the final cleaning step, all pixels with a signal at least 30\% of the core pixels' average and in contact with at least one core pixel are kept. All other camera pixels are removed from the image by setting their signal to zero. 

\begin{SCfigure}[1][tb]
  \centering
  \begin{tabular}[b]{c}
    \includegraphics[width=.31\linewidth]{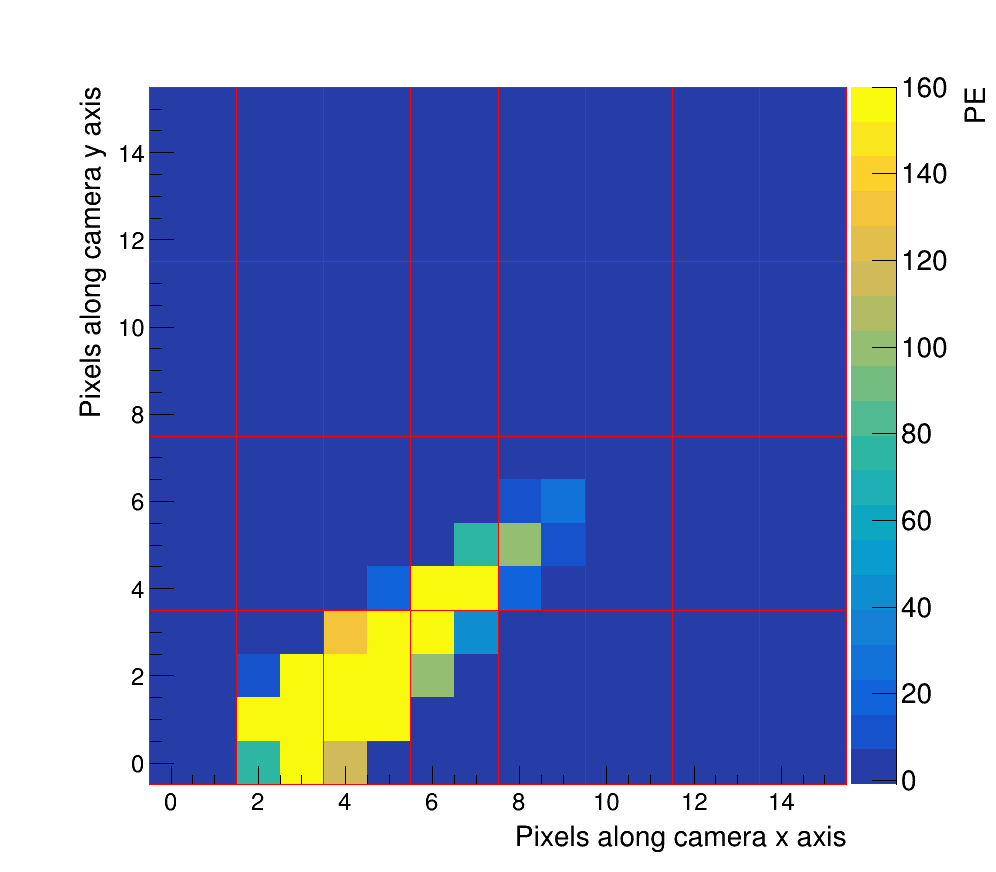} \\
    \small (a)
  \end{tabular} 
  \begin{tabular}[b]{c}
    \includegraphics[width=.31\linewidth]{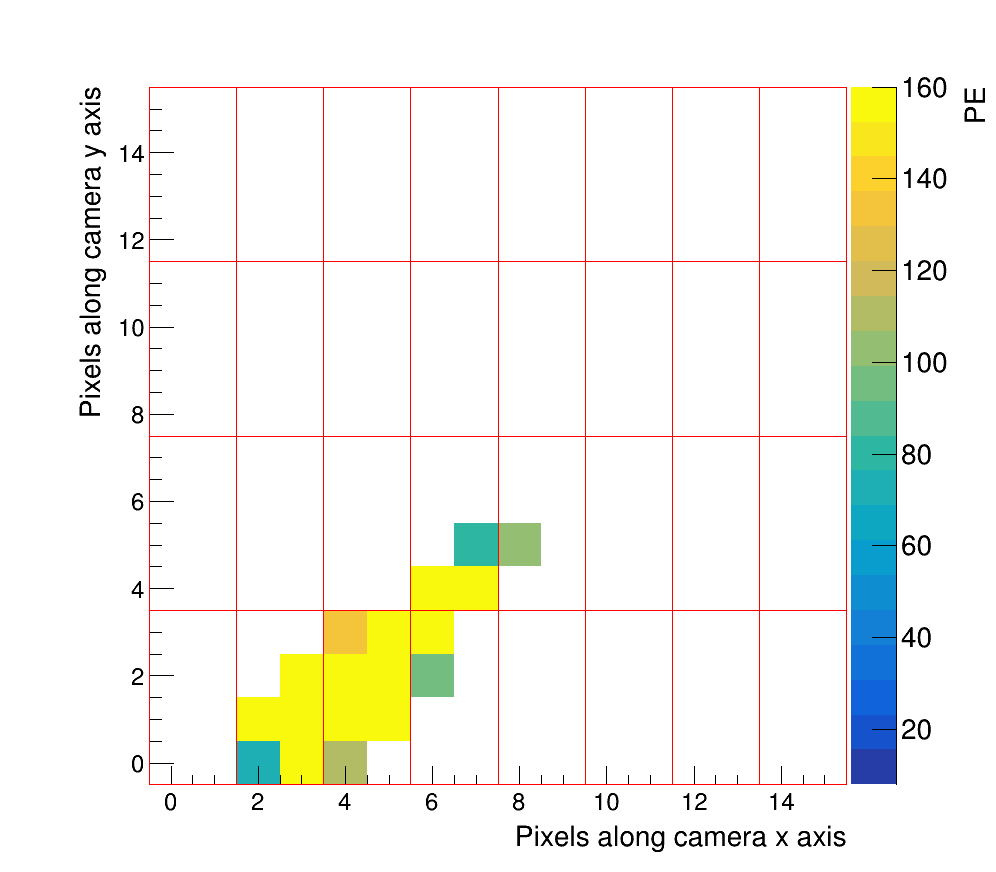} \\
    \small (b)
  \end{tabular}
  \caption{\footnotesize Simulated response of the Demonstrator to a 400\,PeV neutrino induced air-shower originating 170\,km from the telescope and an emergence angle of 1.75$^\circ$. The left panel shows the image before cleaning, and on the right, after cleaning. The color scale gives the per-pixel signal in units of photoelectrons. All figures created using \cite{Philippeal._2025}.}
  \vspace{-2ex}
  \label{fig:cleanedEvent}
\end{SCfigure}

\begin{wrapfigure}[19]{l}{0.6\textwidth}
\includegraphics[trim={2cm 0.8cm 1cm 3.0cm},clip,angle=0,width=0.6\columnwidth]{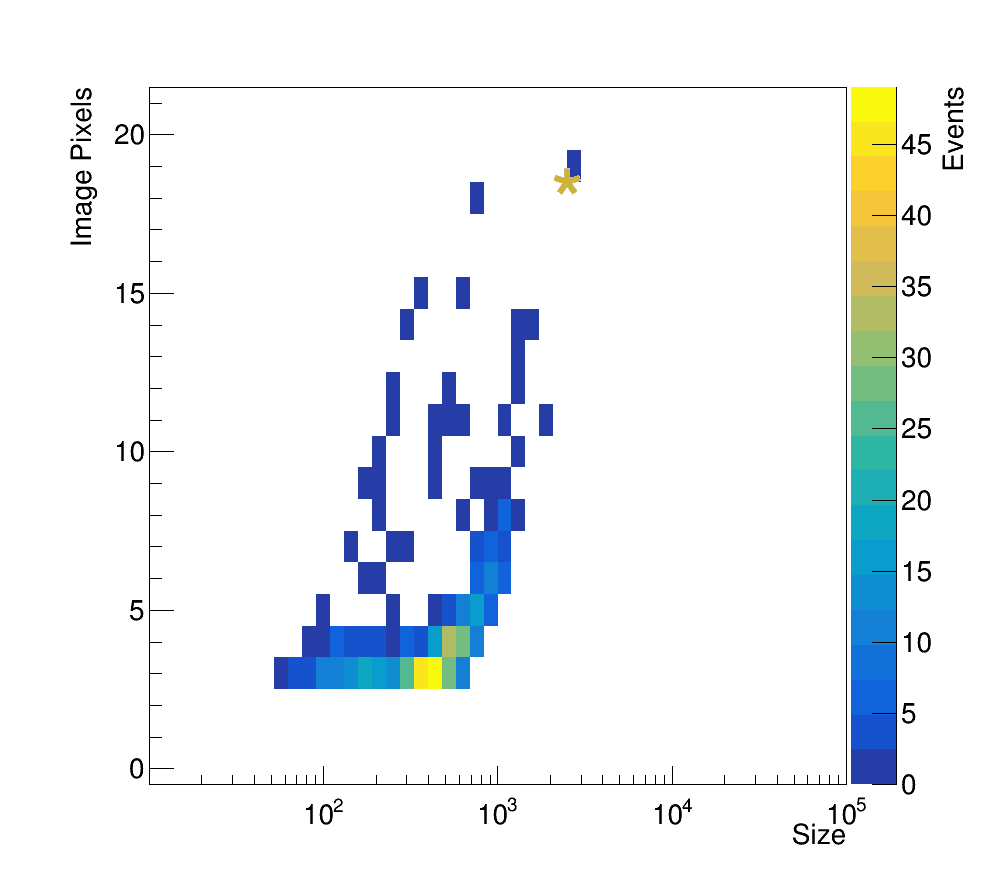}
\caption{\footnotesize Number of pixels in a cleaned event image vs. the \emph{SIZE} of the image. The color scale is the number of events. The star indicates the simulation event parameters shown in figure \ref{fig:cleanedEvent}} 
\label{fig:pixsizedistr}
\end{wrapfigure}

Image cleaning is essential before parameterizing the event with its principal components and reconstructing its energy and arrival direction \cite{Hillas1985a}. Figure \ref{fig:cleanedEvent} shows the cleaned image of a simulated neutrino-induced air shower in the Demonstrator camera. Identifying a similar event in our data would be a gold-plated neutrino candidate event.

Figure \ref{fig:pixsizedistr} shows the distribution of the number of pixels in a cleaned event versus the \emph{SIZE} of an event, \emph{i.e.}, the total number of photoelectrons in the cleaned event. For further analysis, we consider only events with cleaned images of at least three pixels and any \emph{SIZE}. A star marks the position of the gold-plated event from Figure \ref{fig:cleanedEvent} in the pixel vs.\ \emph{SIZE} plane.

\begin{figure}[tb]
  \centering
  \begin{tabular}[b]{c}
    \includegraphics[width=.3\linewidth]{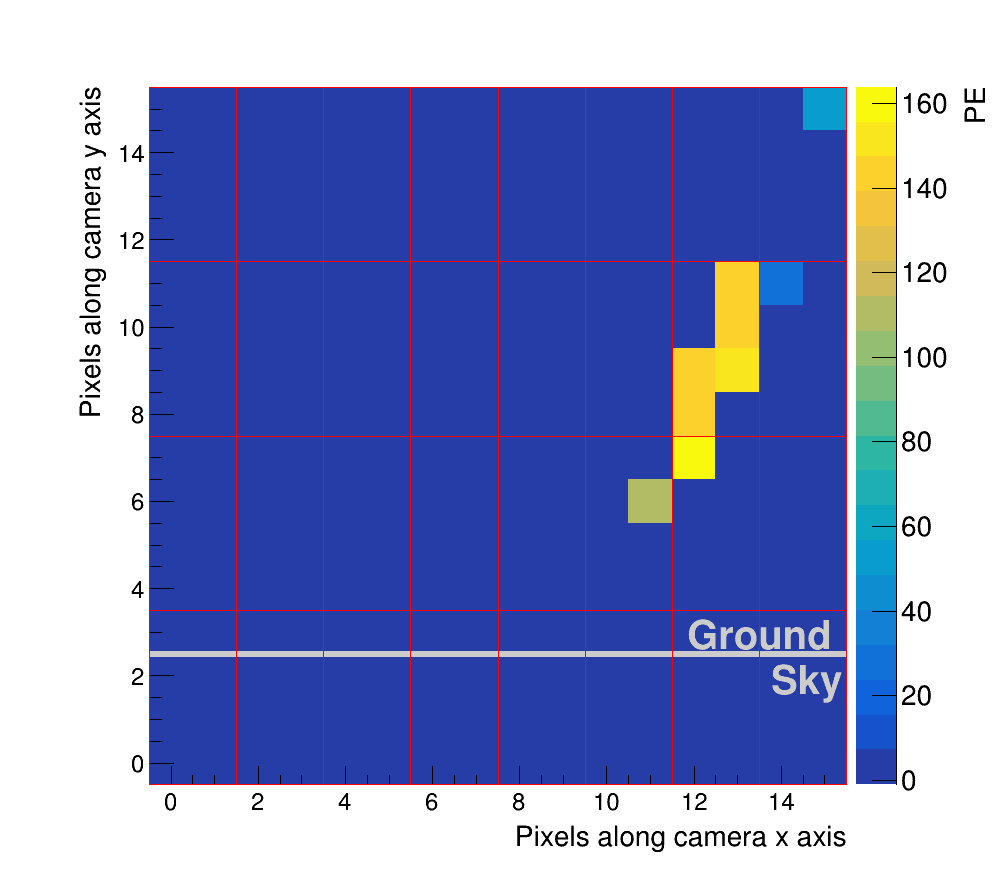} \\
    \small (a)
  \end{tabular} 
  \begin{tabular}[b]{c}
    \includegraphics[width=.3\linewidth]{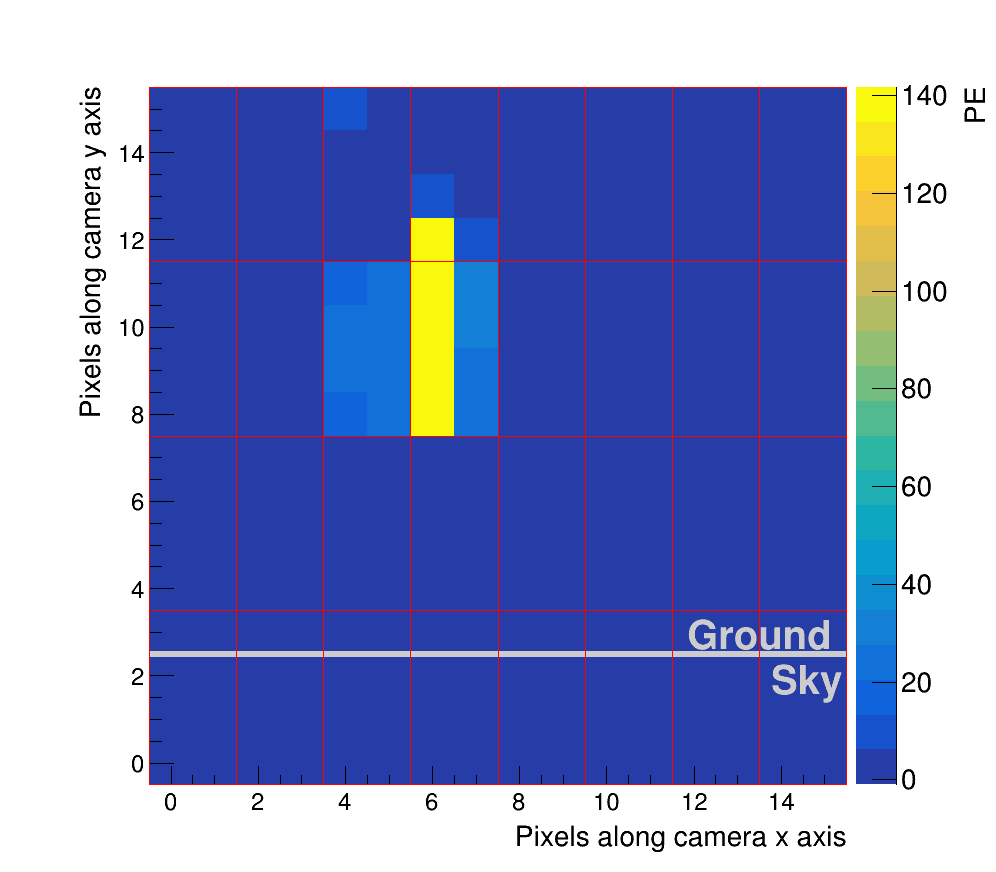} \\
    \small (b)
  \end{tabular}
  \begin{tabular}[b]{c}
    \includegraphics[width=.3\linewidth]{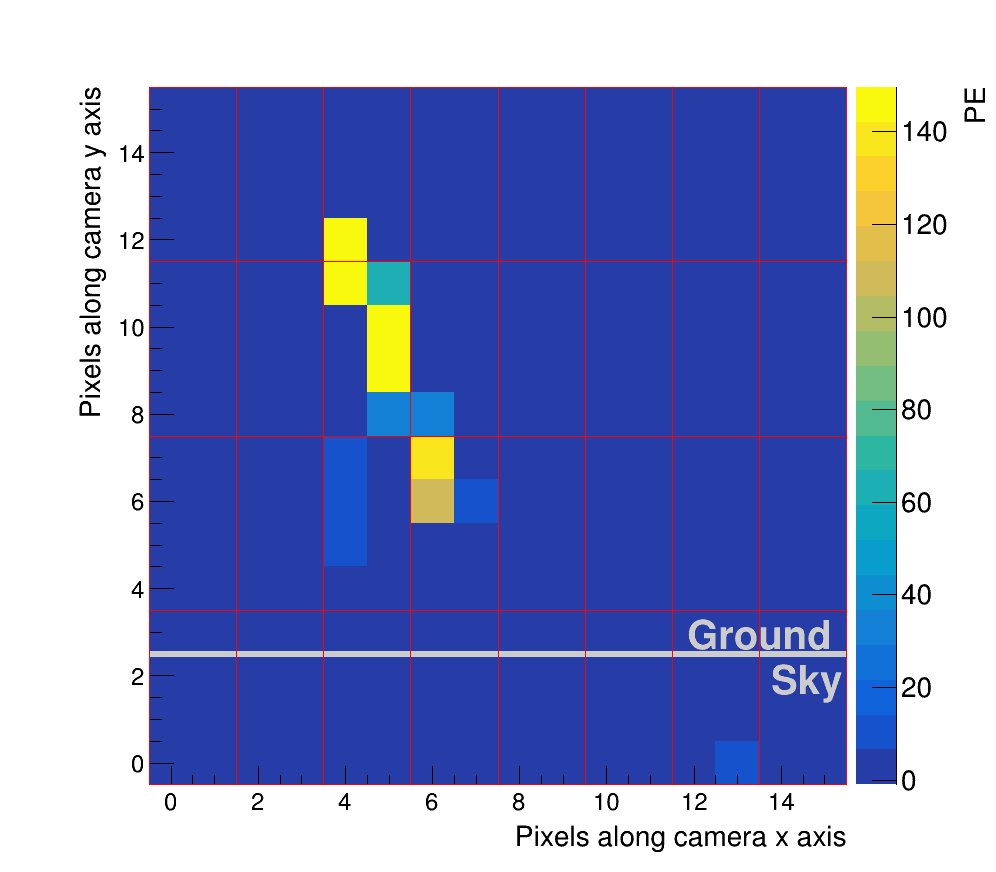} \\
    \small (c)
  \end{tabular}
  \caption{Three sample muon track events from muons traversing the active volume of several SiPMs of the Demonstrator camera. The white horizontal line marks the boundary between the sky and the ground viewed by the camera. The telescope optics flip the position of the sky and ground upside down. The color scale gives the pixel content in photoelectron equivalent signals. \label{fig:muontracks}}
  \vspace{-2ex}
\end{figure}

A total of 497 events survive the image cleaning and have at least three pixels in the cleaned image, \emph{i.e.}, 1.4 events per hour. Visually inspecting the event images, we identify four different event types: clean muon tracks (461 events), muon tracks with signals in parallel SiPMs (15), events that are broad and fuzzy (2), and events caused by a misbehaving camera module (19 events). 

The vast majority, 461 events, are produced by muons traveling through the active volume of several silicon photomultipliers. Figure \ref{fig:muontracks} shows three events with muon tracks. The ionizing trail of the muon, when it traverses the active volume of SiPMs, produces a signal only in the SiPM it passes through. The Demonstrator camera acts as a muon tracker with 3\,mm position resolution in the camera plane (The size of an SiPM is 6\,mm by 6\,mm). To produce a track that fires more than 100 cells in a single SiPM, a muon has to travel through the active SiPM volume with an angle parallel to the SiPM surface of less than $3^\circ$ given that the thickness of the active volume of the SiPM cells is only $\sim100\,\mu$m. Because the acceptance is so small, only 1.4 muon tracks are recorded every hour. Muon tracks are easily identified in the data because they lack the elliptical characteristic of an air shower (see Figure \ref{fig:cleanedEvent}).

\begin{figure}[!htb]
  \centering
  \begin{tabular}[b]{c}
    \includegraphics[width=.3\linewidth]{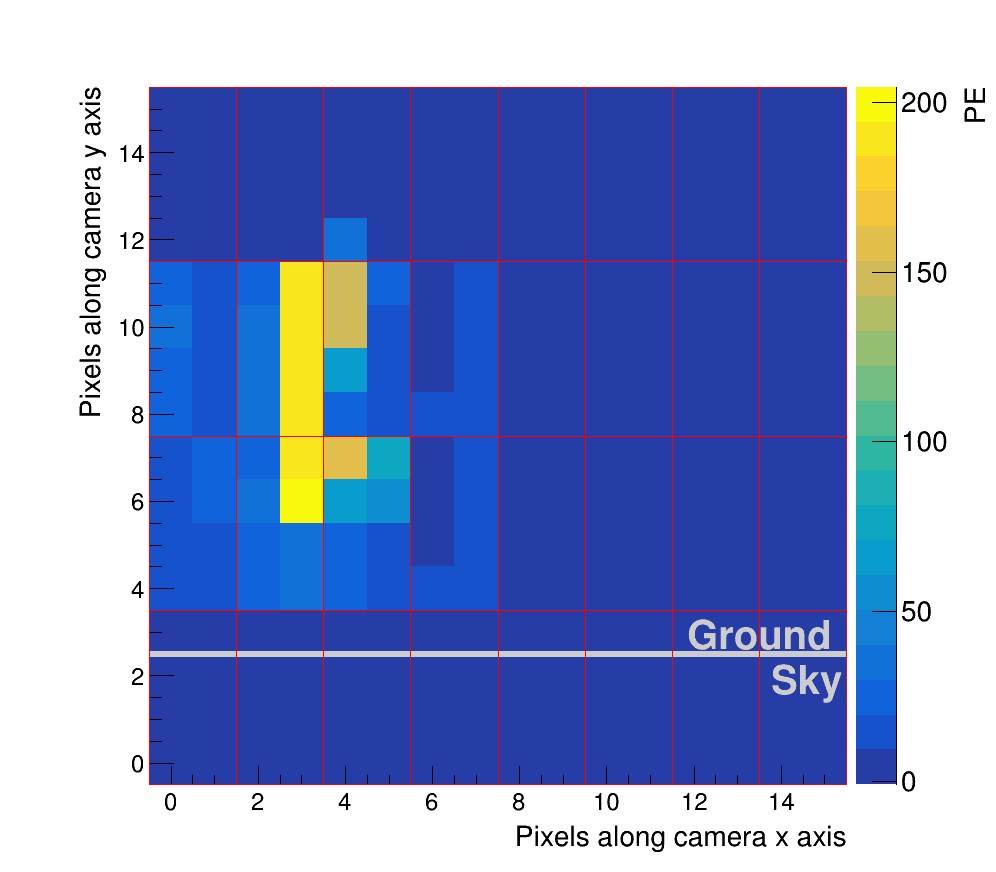} \\
    \small (a)
  \end{tabular} 
  \begin{tabular}[b]{c}
    \includegraphics[width=.3\linewidth]{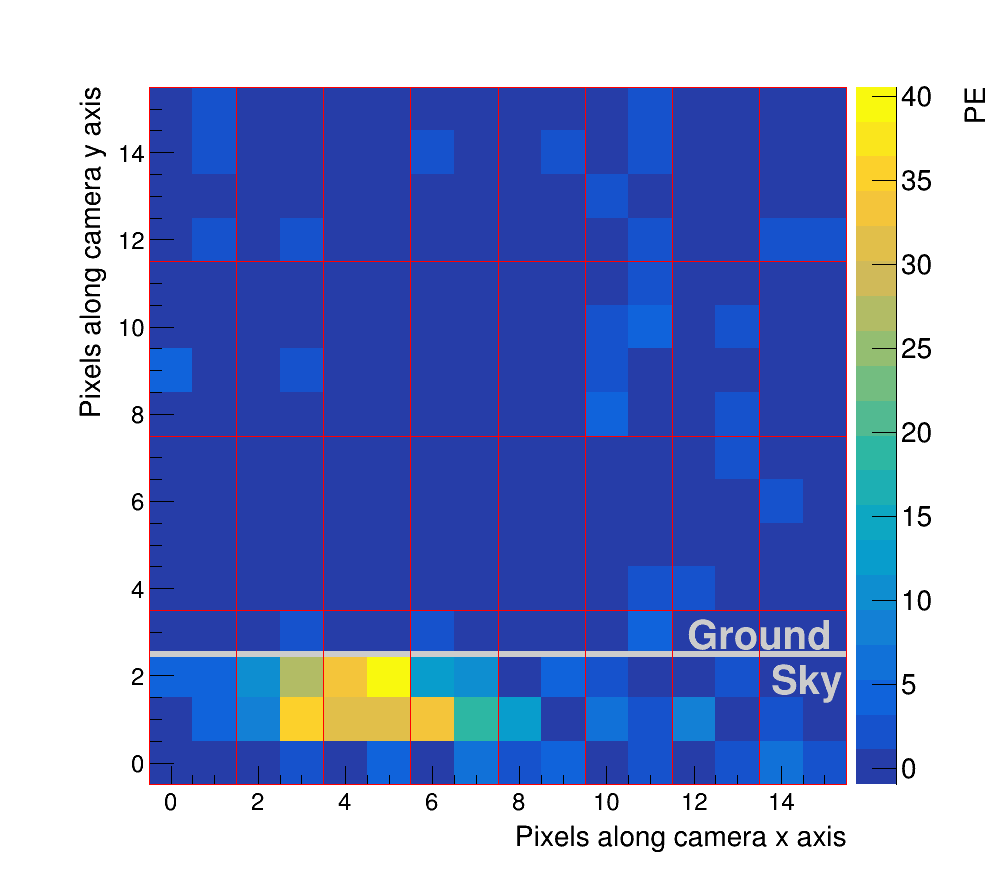} \\
    \small (b)
  \end{tabular}
  \begin{tabular}[b]{c}
    \includegraphics[width=.3\linewidth]{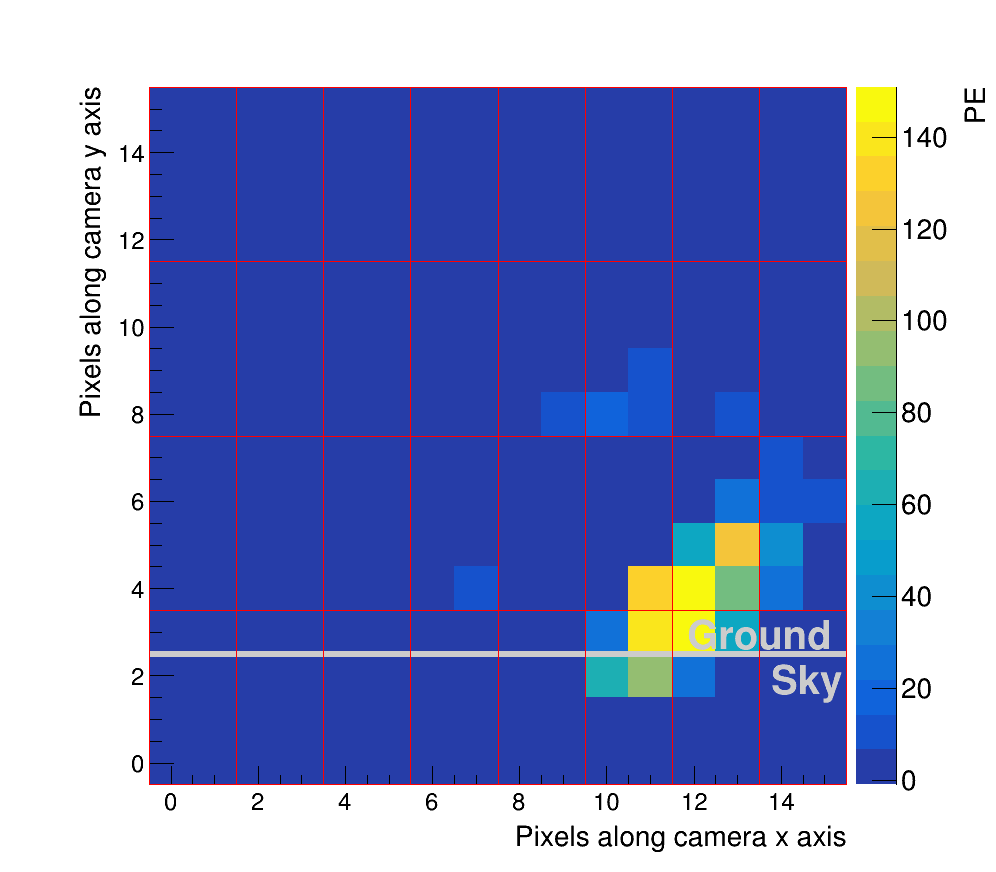} \\
    \small (c)
  \end{tabular}
  \caption{Three sample muon spillover events. The white horizontal line marks the boundary between the sky and the ground viewed by the camera. The telescope optics flip the position of the sky and ground upside down. The color scale gives the pixel content in photoelectron equivalent signals. \label{fig:spillover}}
  \vspace{-2ex}
\end{figure}

We find 15 events ( $1$ event every 23 hours) with signals in SiPMs parallel to the muon track, see Figure \ref{fig:spillover}. These \emph{spillover} events, too, lack the characteristic width of air shower images. However, the signal in the parallel SiPMs cannot be explained by one single muon traveling along a straight line. A potential explanation is delta electrons produced in a muon-electron interaction and propagating into a neighboring pixels.
\begin{figure}[!tb]
  \centering
  \begin{tabular}[b]{c}
    \includegraphics[width=.4\linewidth]{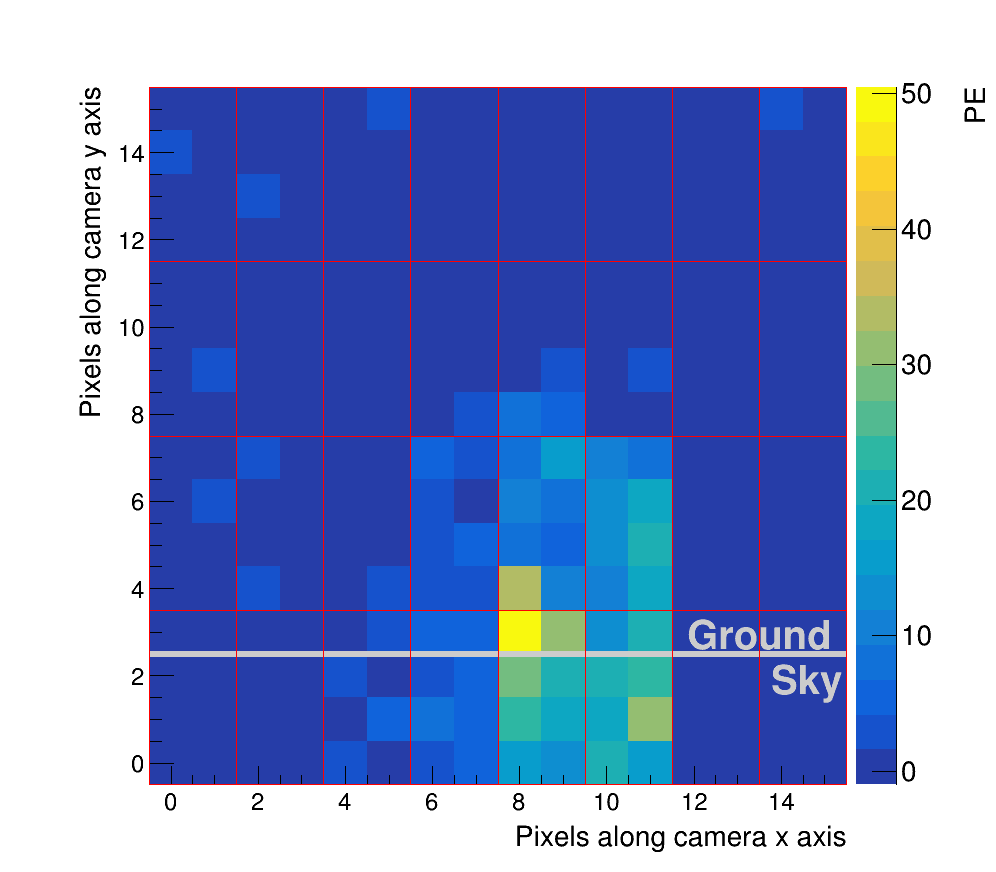} \\
    \small (a)
  \end{tabular} 
  \begin{tabular}[b]{c}
    \includegraphics[width=.4\linewidth]{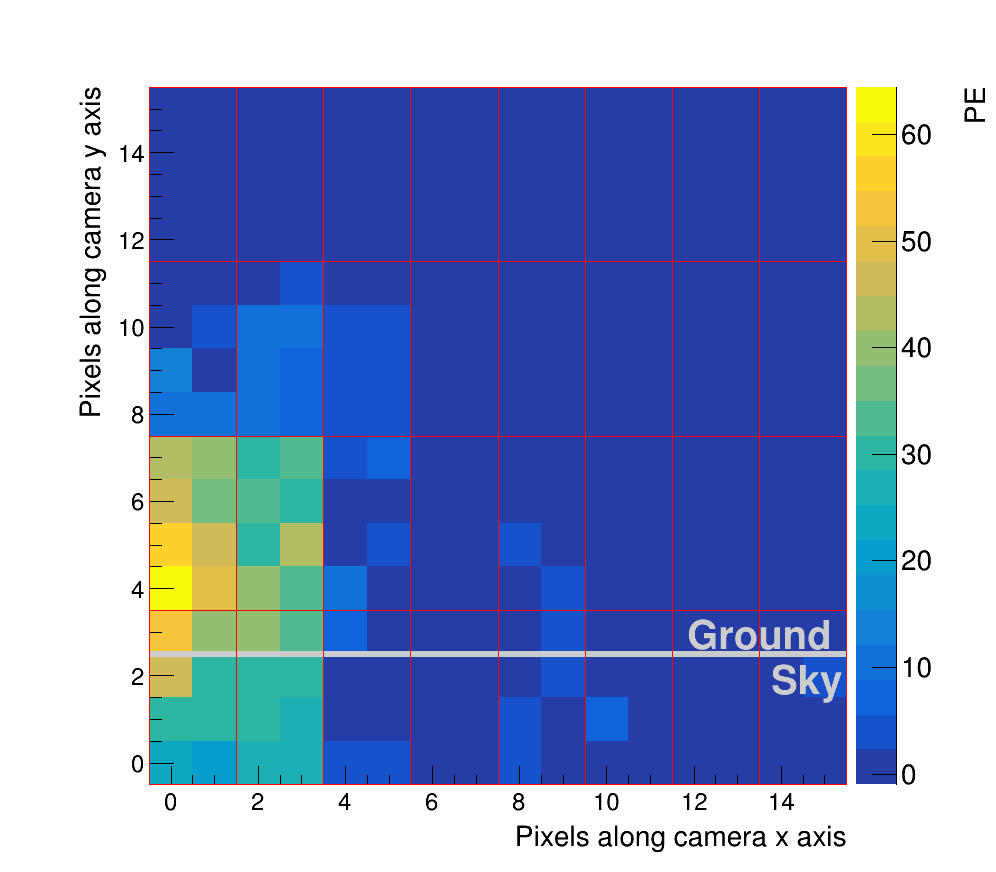} \\
    \small (b)
  \end{tabular}
  \caption{The two fuzzy events. The white horizontal line marks the boundary between the sky and the ground viewed by the camera. The telescope optics flip the position of the sky and ground upside down. The color scale gives the pixel content in photoelectron equivalent signals.\label{fig:fuzzy}}
  \vspace{-2ex}
\end{figure}

Figure \ref{fig:fuzzy} shows the two events in our event selection that do not appear like muon tracks, although a charged particle interaction cannot be excluded. These \emph{fuzzy }events are characterized by several bright pixels within a $1^\circ$ radius, forming an image that is dissimilar to the expected well-defined air-shower image in Figure \ref{fig:cleanedEvent}. We thus do not consider these to be valid air shower images, either for now, but we will continue to investigate their nature.\\

\begin{wrapfigure}[17]{l}{0.4\textwidth}
  \centering  
  \includegraphics[width=0.4\columnwidth]{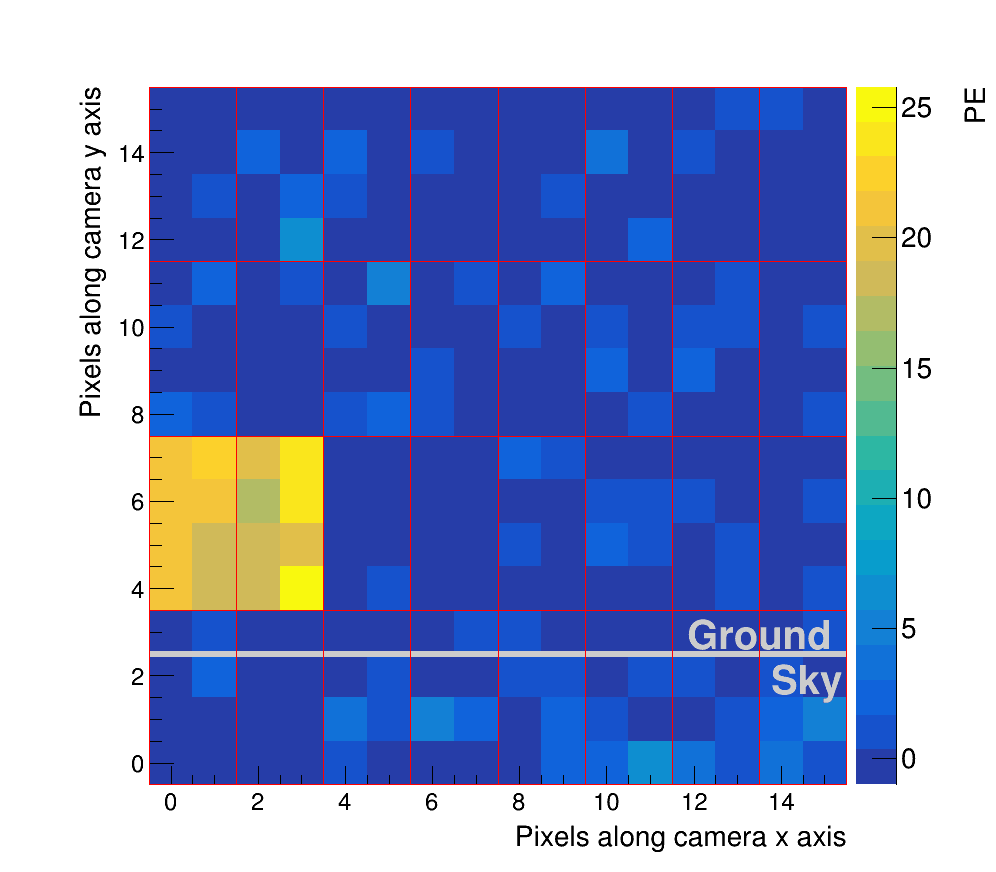} \\
  \caption{A sample event caused by a misbehaving camera module. The white horizontal line marks the boundary between the sky and the ground viewed by the camera. The telescope optics flip the position of the sky and ground upside down. The color scale gives the pixel content in photoelectron equivalent signals.\label{fig:BadModule}}
\end{wrapfigure}

Nineteen events can be traced back to a temporarily misbehaving camera module on four different days, lighting up all pixels of the module; see Figure \ref{fig:BadModule}. Given their unique topology, these events can be unambiguously removed from the analysis.
\vspace{-4mm}
\section{Discussion}

The \emph{Trinity} Neutrino Observatory targets the PeV to EeV energy range. Since Fall 2023, we have operated the \emph{Trinity} Demonstrator. Here, we present the results of analyzing 346 hours of Demonstrator data to identify background sources for \emph{Trinity}. We find that charged particle interactions with the SiPMs, \emph{i.e.}, direct muon hits, leaving signals in several SiPMs are the main source of background. The Demonstrator records about two of these line-like events every hour. Muon events can be easily identified in the data as they do not mimic a gold-plated neutrino shower event. Furthermore, these muon events can be rejected during data taking in the future with a plastic-scintillator-based veto detector surrounding the camera focal plane.

Given that a qualitative analysis removes all but two events in our 346-hour observation, it demonstrates the resiliency of the \emph{Trinity} approach. The remaining two events (1 per 200 hours) do not look like the gold-plated event and are most likely rejected in a quantitative analysis based on an end-to-end simulation chain we are currently commissioning. The results demonstrate that we can sufficiently suppress the background in the data, and a muon veto will reliably eliminate all muon events during data taking. 

Given the Demonstrator results, we are ready to embark on the next phase towards the \emph{Trinity} Neutrino Observatory, which will be \emph{Trinity} One. \emph{Trinity} One will be the first complete \emph{Trinity} telescope, that can rotate in azimuth. We project \emph{Trinity} One to have an excellent point-source sensitivity, which we discuss elsewhere in these proceedings \cite{Raudales_2025}.\\
\textbf{This work was supported with NSF awards PHY~2112769 and PHY~2411666}





\begingroup
\setstretch{0.2}
\bibliographystyle{apsrev4-2}
\bibliography{main}
\endgroup

\pagebreak

\section*{Author List}

\noindent\includegraphics[width=0.35\linewidth]{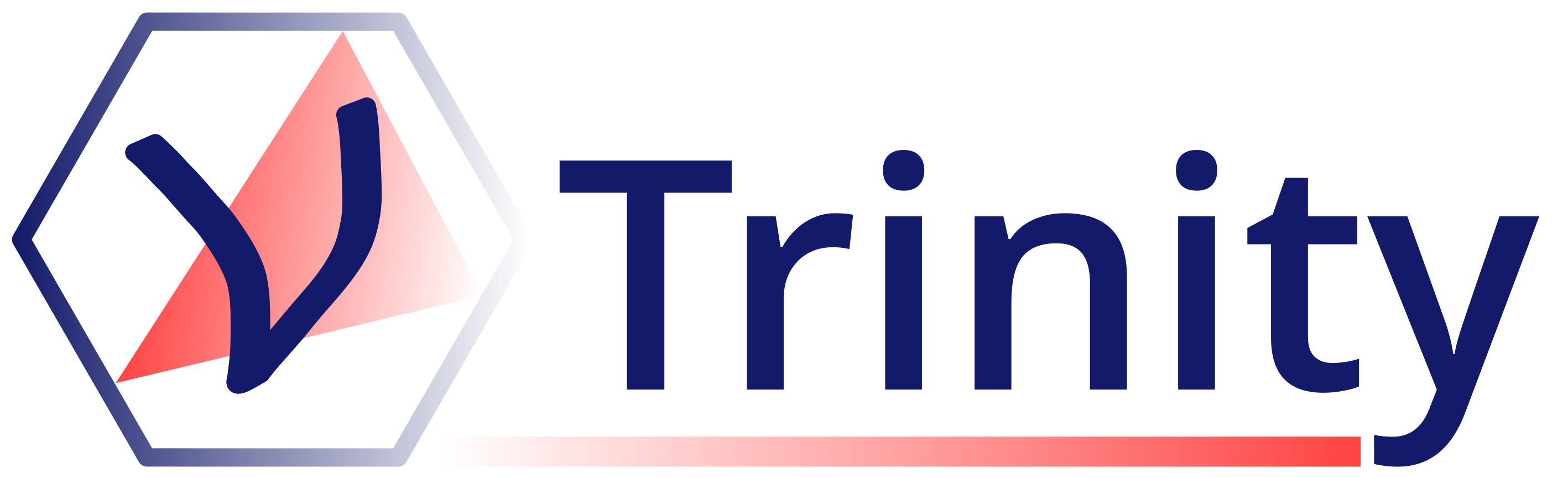}
\vspace{0.5em}

\small
M. Bagheri$^{1}$,
A. Barletta$^{1}$,
D. Bergmann$^{2}$,
J. Blose$^{1}$,
J. Bogdan$^{1}$,
A. M. Brown$^{3}$,
L. Cedeno$^{1}$,
M. Doro$^{4}$,
M. Fedkevych$^{1}$,
S. Gadamsetty$^{1}$,
F. Giordano$^{5}$,
C. Hao$^{1}$,
V. Iyengar$^{1}$,
D. Kieda$^{2}$,
N. Lew$^{1}$,
M. Mariotti$^{3}$,
Y. Onel$^{6}$,
A. N. Otte$^{1}$,
D. A. Raudales O.$^{1}$,
L. Rojas Castillo$^{1}$,
A. Ronemus$^{1}$,
A. Menon$^{1}$,
A. Mitra$^{1}$,
E. Schapera$^{1}$,
D. Solden$^{2}$,
N. Song$^{1}$,
W. Springer$^{2}$,
S. Stepanoff$^{1}$,
I. Taboada$^{1}$,
K. Tran$^{2}$,
A. Wilcox$^{1}$,
A. Zhang$^{1}$

\medskip

\noindent
$^{1}$School of Physics and Center for Relativistic Astrophysics, Georgia Institute of Technology, Atlanta, GA, USA \\
$^{2}$Department of Physics and Astronomy, University of Utah, Salt Lake City, UT, USA \\
$^{3}$Department of Physics and Centre for Advanced Instrumentation, Durham University, Durham, UK \\
$^{4}$INFN Sezione di Padova and Università degli Studi di Padova, Padova, Italy \\
$^{5}$INFN Sezione di Bari and Università degli Studi di Bari, Bari, Italy \\
$^{6}$Physics and Astronomy Department, The University of Iowa, Iowa City, USA \\
\end{document}